\documentclass{aa}
\usepackage{txfonts}
\usepackage{natbib}
\usepackage{graphicx}
\usepackage{aalongtable}
\bibpunct{(}{)}{;}{a}{}{,}
\begin{document}

\title{Spectroscopy of horizontal branch stars in \object{NGC6752}}
\subtitle{Anomalous results on atmospheric parameters and masses
\thanks{Based on observations with the ESO Very Large Telescope
at Paranal Observatory, Chile (proposal ID 69.D-0682)}
}

\author{
C. Moni Bidin \inst{1}\fnmsep\inst{2}
\and
S. Moehler \inst{3}
\and
G. Piotto \inst{1}
\and
Y. Momany \inst{4}
\and
A. Recio-Blanco \inst{5}
}

\institute{
Dipartimento di Astronomia, Universit\`{a} di Padova,
Vicolo dell'osservatorio 3, 35122 Padova, Italy
\and
Departamento de Astronom\'{i}a, Universidad de Chile,
Casilla 36-D, Santiago, Chile
\and
European Southern Observatory, Karl-Schwarzschild-Str. 2,
85748 Garching, Germany
\and
INAF-Osservatorio Astronomico di Padova, 
Vicolo dell'osservatorio 2, 35122 Padova, Italy
\and
Observatoire de la C\^{o}te d'Azur,
Dpt. Cassiop\'ee, CNRS UMR 6202,
B.P. 4229, 06304 Nice, Cedex 04, France
}
\date{Received / Accepted }


\abstract{The determination of fundamental parameters for horizontal
hranch stars in Galactic globular clusters frequently gives puzzling
results, pointing to the lack of our understanding of their
atmospheric structure and the inadequate approximations by the
models.}  {We wanted to measure effective temperatures, surface
gravities, helium abundances, and masses for hot horizontal branch
stars in \object{NGC6752} in order to compare the results with
evolutionary predictions.}  {We used the ESO VLT-FORS2 facility to
collect low-resolution spectra of 51 targets distributed along the
horizontal branch. We determined atmospheric parameters, by comparison
with theoretical models through standard fitting routines, and masses
by basic equations.}  {Results generally agree with
previous works, although not always with the theoretical expectations for
cooler stars (T$_\mathrm{eff}\leq$15\,000~K).  The calculated color
excess is systematically lower than the literature values, pointing
towards a possible underestimation of effective
temperatures. Moreover, we find two groups of stars at
T$_\mathrm{eff}\sim14\,000$~K and at T$_\mathrm{eff}\sim27\,000$~K that
present anomalies with respect to the general trend and expectations.
We suppose that the three peculiar bright stars at
T$_\mathrm{eff}\sim14\,000$~K are probably affected by an enhanced
stellar wind. For the eight extreme horizontal branch stars at
T$_\mathrm{eff}\sim27\,000$~K that show unusually high masses, we find
no plausible explanation.}  {While most of our results agree well with
the predictions of standard horizontal branch evolution, we still have
problems with the low masses we derive in certain temperature
ranges. We believe that Kurucz ATLAS9 LTE model atmospheres with
solar-scaled abundances are probably inadequate for these temperature
ranges. Concerning the group of anomalous stars at
T$_\mathrm{eff}\sim27\,000$~K, a Kolmogorov-Smirnov test indicates that
there is only an 8.4\% probability that these stars are randomly drawn
from the general distribution in the color-magnitude diagram. This is
not conclusive but indicates that these stars could be both (and
independently) spectroscopically and photometrically peculiar with
respect to the general extreme horizontal branch population.}

\authorrunning{Moni Bidin et al.}
\maketitle


\section{Introduction}
\label{capintro}

Many decades have passed since when horizontal branch (HB) stars
were identified as evolved stars of low initial mass (approximately
0.7-2 M$_{\sun}$) presently burning helium in their core
\citep{HoyESchwarz55,Faulk66}. Still, many aspects of the
evolution and internal structure of these stars are not fully
understood.  In particular, in the color-magnitude diagrams of
Galactic globular clusters, many well-established observed features of
HB stars are good example of this, as seen in the case of the different blue extensions
from one cluster to another, which are partially dependent on
metallicity \citep{Sandage60} but not fully explained by it
\citep{Sandage67,VanDenBergh67}. In addition, the HB
may show ``jumps'', where stars appear brighter than the theoretical
expectations \citep{Grundahl99,Momany02,Momany04}, or gaps, i.e. underpopulated
regions along the HB \citep{Sosin97,Catelan98,Ferraro98,Piotto99}.
Spectroscopy of HB stars opened new questions, such as the unexplained presence of
fast rotators among stars redder than the \citet{Grundahl99} jump
\citep{Peterson83,Peterson85a,Peterson85b,Behr03,Behr00,Behr00b, Recio04}
and the absence of close binaries among hot extreme horizontal branch
(EHB) stars \citep{Moni06b,Moni06a}.

Traditionally, low-resolution spectroscopy of HB stars has focused on
determining of their fundamental quantities such as atmospheric
parameters.  This is usually achieved by comparing observed spectra with
grids of theoretical ones obtained from model atmospheres. As a
consequence, any aspect of stellar structure unaccounted for in the
model may be directly reflected on unexpected values measured for the
parameters.  For example, it is well known that surface gravities are
systematically lower than expected in the temperature range 12\,000-20\,000~K
if model spectra with the metallicity of the parent globular
cluster are used. As a consequence, masses are derived that are too low
\citep{Moehler95}. This ``mass discrepancy'' can be partially
explained by radiative levitation of heavy elements \citep{Grundahl99},
originally proposed as the physical process resposible for abundance anomalies
observed in field Ap stars \citep{Michaud70}.
Precise calculations confirmed that this phenomenon should be at work in the
atmospheres of these stars and can account for observed anomalies
\citep{Michaud83}, but
the problem of the mass discrepancy is still not completely solved
\citep{Moehler01}.
On the other hand, \citet{Vink02}
point out that neglecting the presence of stellar wind can cause
measured surface gravities to be erroneously low.
Nevertheless,
the derived values of stellar parameters - even if erroneous or
unexpected - have been useful for pointing out where canonical models are
inadequate, stimulating further investigations to solve the open problems.

As part of a program aimed at searching for close binaries among HB stars
\citep[][ hereafter Paper I]{Moni06a}, we also collected low-resolution
spectra of our target stars. In this paper we
present the analysis of these spectra and critically discuss the
derived atmospheric parameters (effective temperature, surface
gravity, helium abundance) and stellar masses.

\section{Observations and data reduction}
\label{capdata}

\begin{figure}
\begin{center}
\resizebox{\hsize}{!}{\includegraphics{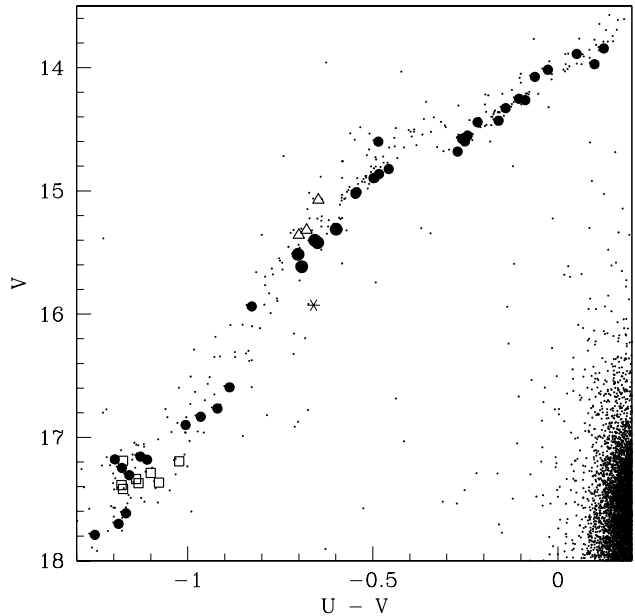}}
\caption{
Observed stars in the $V$ vs $U-V$ color-magnitude diagram. Photometric data
are from \citet{Momany02}. The stars are indicated with different symbols,
as discussed in the text. The same symbols will be used in all of
the figures to distinguish groups of stars with different properties.}
\label{cmd}
\end{center}
\end{figure}

The spectra were collected during two nights of observations (June 12
and 13, 2002) at the VLT-UT4 telescope equipped with the spectrograph
FORS2 in MXU mode.  The fifty-one target stars are the same objects
already described in Paper I.  They are distributed along the entire HB,
spanning a wide range in temperature from $\mathrm{T_{eff}} \approx 8\,000 K$
up to $\mathrm{T_{eff}} \approx 30\,000 K$.  Coordinates and
photometric data of the observed stars are presented in Table 1 of
Paper I.
Figure \ref{cmd} shows the position of
target stars in the color-magnitude diagram.

\begin{figure*}
\begin{center}
\includegraphics[width=17cm]{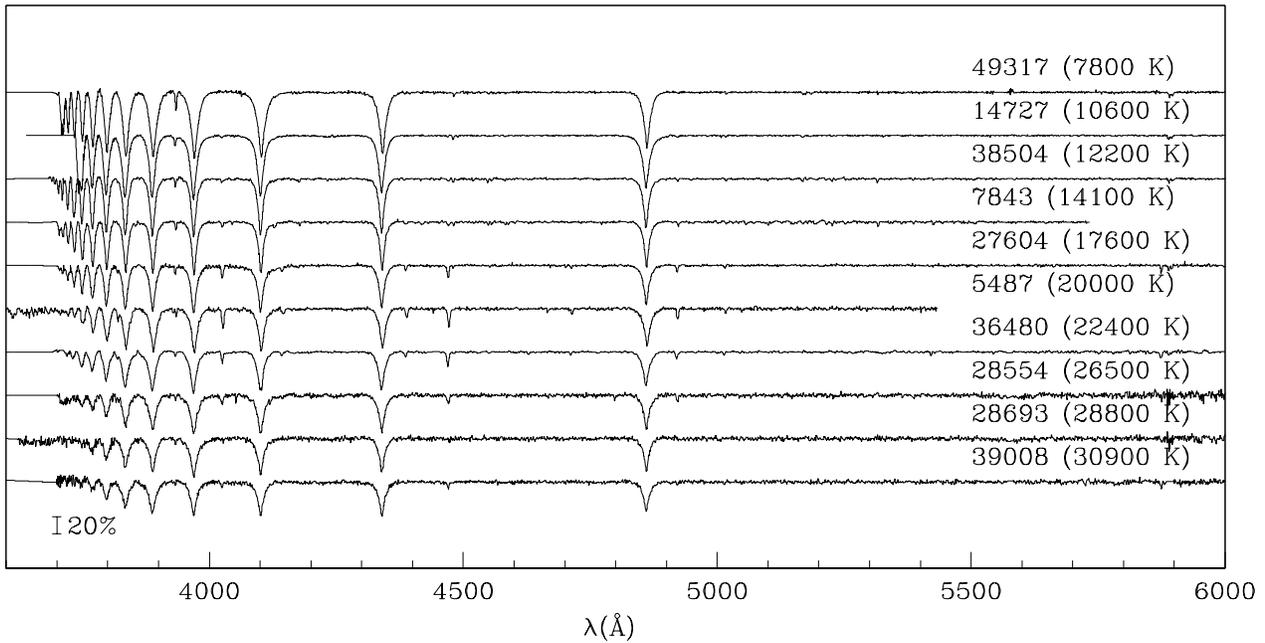}
\caption{A sample of the collected spectra,
normalized and ordered by increasing $\mathrm{T_{eff}}$ from
top to bottom.
Spectra were normalized fitting a 3$^{rd}$- order spline function to the
continuum. For wavelengths shorter than 3900 \AA, where Balmer lines blend,
the fit was anchored at the highest points between lines.
}
\label{spectrafig}
\end{center}
\end{figure*}

Two 1350s exposures were secured for each target with grism 600B+22
and 0\farcs5 -- wide slits for a resulting resolution of 3 \AA.
The spectral range was approximately 2900 \AA\ wide and, on average, centered
at 4600 \AA, but slightly different from star to star due to the different
positions of the slits in the mask used for multi-object
spectroscopy. All Balmer lines from H$_{\beta}$ to H10 were always
present in the spectra.  We never used wavelengths shorter than 3600 \AA
in our analysis, because of the lack of atmospheric transmission and
instrumental response.  In Fig. \ref{spectrafig} we show an
example of the reduced spectra.

The calibration images (bias, flat, and lamp) were acquired before and
at the end of each observing night. Data reduction was performed with
standard MIDAS\footnote{ ESO-MIDAS is the acronym for the European
Southern Observatory Munich Image Data Analysis System which is
developed and maintained by the European Southern Observatory
(http://www.eso.org/projects/esomidas/)} procedures.  All of the 2D
spectra were trimmed from the frames with their calibration
images and reduced independently.  The wavelength calibration was
performed with the HeHgCd lamp images, fitting a $3^\mathrm{rd}$
order polynomial to the dispersion relation.
The rms error indicated by the fitting procedure was 0.07 \AA on average.
Before extraction, we rebinned the 2D images to constant
wavelength steps of 0.4 \AA /pix and corrected their curvature along
the spatial axis by tracing the spectra as described in
\citet{Moehler06}.  Spectra were then extracted both with an optimum
extraction algorithm \citep{Horne86} and with a simple sum, and then,
for each star, we chose the procedure that gave better results
in terms of S/N, as the optimum algorithm sometimes failed
to correctly extract the spectra of bright stars.
We corrected them for atmospheric extinction, with the extintion
coefficients for the La Silla observatory \citep{Tug77}.
The response curve was obtained
with observations of the standard stars \object{EG274} (both nights)
and \object{LTT3218} (second night only),
using the flux tables of \citet{Hamuy94}. We had to use a high-order
polynomial (6th) so as to obtain a good response curve and a smooth
continuum on the calibrated spectra.
The resulting curves had the same shape with no difference between the nights.
To calibrate the spectra of the second night we averaged the two corresponding
curves when two standard stars were observed.

Finally, we corrected the spectra for radial velocity shifts.
Absolute radial velocities of the targets were already measured on higher resolution
spectra as described in Paper I, but we measured them again
on each of the low-resolution spectra used in this work.
In fact, as already discussed in Paper I, the shift of the spectra
with respect to laboratory
wavelength was due not only to the radial velocities of the stars, but also to
many effects such as the position of the stars in the slitlets
and systematics in wavelength calibration.
We fitted the cores of all Balmer lines from H$_{\beta}$ to
H$_{9}$ with a Gaussian profile, excluding H$_\epsilon$ due to blending with
the \ion{Ca}{ii}~H line, and assumed the average as the radial
velocity to be corrected.

Usually the values from single lines agreed within 10 km s$^{-1}$, but
in some cases there were large differences (up to 50 km s$^{-1}$),
an effect already described in \citet{Moehler97}.
We checked if these could be ascribed at least in part to a rigid shift
of the spectra on the CCD, constant in pixel and then variable with wavelength in km s$^{-1}$,
but we found no evidence of a clear trend with wavelength.


\section{Measurements}
\label{capmeasure}

\begin{figure}
\begin{center}
\resizebox{\hsize}{!}{\includegraphics{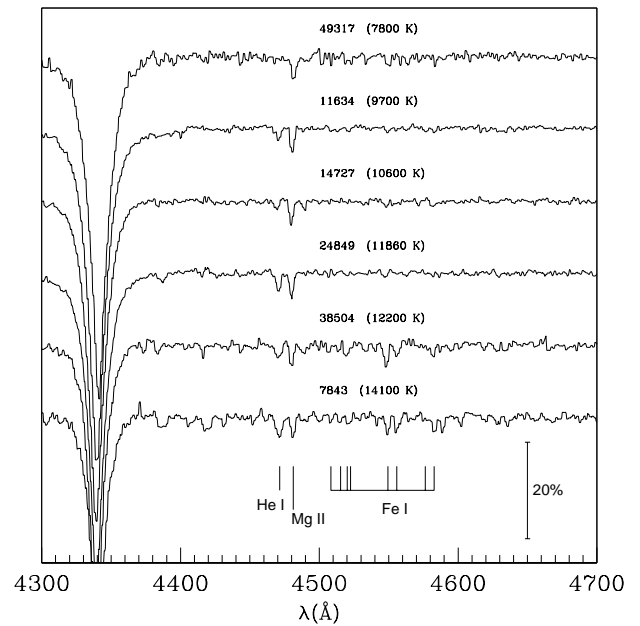}}
\caption{A sample of spectra for stars across the Grundhal jump in the spectral
range used to determine the metallicity of the model in the fitting
procedure. Some of the most relevant \ion{Fe}{ii} are indicated with the
\ion{Mg}{ii} and \ion{He}{i} doublets that are the most important features in this spectral range.} 
\label{figFeregion}
\end{center}
\end{figure}

In order to derive effective temperatures, surface gravities, and helium abundances,
we fitted the observed Balmer and Helium lines with stellar model
atmospheres.

As discussed extensively in recent years, HB stars hotter than about
11\,000 -- 12\,000~K show evidence of
diffusion, e.g.  deficiency of helium and strong (solar to super-solar levels)
enrichment in iron (\citealt{Grundahl99,Glaspey85},
\object{NGC6397}; \citealt{Glasp89,Moehler00}, \object{NGC6752};
\citealt{Behr99,Moehler03}, \object{M13}; \citealt{Behr00}, \object{M15};
\citealt{Fabbian05}, \object{NGC1904}; \citealt{Pace06}, \object{NGC2808}).
As the atmospheric metallicity has a non-negligible influence on the
profiles of hydrogen and helium lines, we need to know if diffusion is
active in a star to decide on the metallicity to fit our spectra.
Therefore, in all spectra we carefully examined the region
4450~\AA\ to 4600~\AA, where strong \ion{Fe}{ii} lines show up at
11\,000~K to 12\,000~K. Past experience \citep{Moehler99,Moehler00,Moehler03}
has shown that these lines are strong enough to allow detection even in
medium-resolution spectra.
In Fig. \ref{figFeregion} we show this wavelength range for some spectra
of stars across the Grundhal jump.
Spectra showing evidence of iron lines
{\em or} of being hotter than 14\,000~K (deduced from their position
in the color-magnitude diagram) were fit with metal-rich ([M/H] $+0.5$)
model spectra, whereas stars below 14\,000~K with no evidence of strong
iron lines were fit with metal-poor model
[M/H] = $-1.5$) spectra.
In the cool star spectra, we kept the helium
abundance fixed, as the helium lines in cool stars are rather weak, and
the helium abundance should be close to solar anyway.
During the fitting we verified that those helium lines predicted for
these cool stars agree with the observed ones.
We computed model
atmospheres using ATLAS9 \citep{Kurucz93} and used Lemke's
version\footnote{For a description see
http://a400.sternwarte.uni-erlangen.de/$\sim$ai26/linfit/linfor.html}
of the LINFOR program (developed originally by Holweger, Steffen, and
Steenbock at Kiel University) to compute a grid of theoretical spectra
that include the Balmer lines H$_\alpha$ to H$_{22}$, \ion{He}{i}, and
\ion{He}{ii} lines. The grid covered the range
7\,000~K~$\leq$~T$_\mathrm{eff}$~$\leq$~35\,000~K,
2.5~$\leq$~$\log{g}$~$\leq$~6.0,
$-3.0$~$\leq$~$\log{He}$~$\leq$~$-1.0$, at metallicities of
[M/H]~=~$-1.5$ and $+$0.5. In Table \ref{datatable} we
list the results obtained from fitting the Balmer lines H$_\beta$ to
H$_{10}$ (excluding H$_\epsilon$ to avoid the \ion{Ca}{ii}~H line)
and, in the hot stars, also the \ion{He}{i} lines 4026~\AA, 4388~\AA,
4471~\AA, 4921~\AA\ and the \ion{He}{ii} lines 4542~\AA\ and 4686~\AA.
IDs are from \citet{Momany02}. In cool star spectra the He abundance was kept fixed
(-1.00$\pm$0.00).
Errors comes from the weighting procedure of two measurement for each star,
and are multiplied by $\sqrt{3}$ but not corrected for
possible underestimation. Masses were derived with Eq. \ref{eq1}.
In last column peculiar stars are indicated, with the same symbols used in all figures.

To establish the best fit to the observed spectra, we used the routines
developed by \citet{Bergeron92} and \citet{Saffer94},
as modified by \citet{Napiwotzki99},
which employ a $\chi^2$ test. The $\sigma$ necessary for the
calculation of $\chi^2$ is estimated from the noise in the continuum
regions of the spectra. The fit program normalizes model spectra {\em
and} observed spectra using the same points for the continuum
definition.

For each star, the atmospheric parameters were measured
independently in the two low-resolution spectra collected during the
observing run. The final results given in Table \ref{datatable} are
the weighted means of these values, weighted by the inverse
errors provided by the fit procedure.
The tabulated errors are the
results of the weighted mean procedure applied to the pair of
independent measurements obtained by the fit of the two spectra for each star.
We multiplied the errors given by the fitting routine by $\sqrt{3}$
because, when rebinning, we oversampled the spectra by a factor of three
with respect to the dispersion, while the fitting procedure
assumes each pixel is independent of the others.
The errors in each fit of the observed spectrum with the model
spectrum were derived from the $\chi^2$ of the fit itself
\citep[see][for more details]{Moehler99}. These errors are obtained under
the assumption that the only error source is the statistical noise
(as derived from the continuum of the spectrum). However, Napiwotzki
(priv. comm.) noted that the routine underestimates this statistical error
by a factor of 2-4.

\begin{table}[t!]
\begin{center}
\caption{Atmospheric parameters and derived masses for target stars.}
\label{datatable}
\begin{tabular*}{9cm}[t]{ c r c c c l }
\hline \hline
ID&$\mathrm{T_{eff}}$ (K)&$\log{g}$&$\log{\frac{N(He)}{N(H)}}$&M (M$_{\sun}$)\\
\hline
14770&28400$\pm$300&5.53$\pm$0.03&$-$2.25$\pm$0.05&0.55$\pm$0.04 &\\
11634&9700$\pm$100&3.39$\pm$0.05&$-$1.00$\pm$0.00&0.39$\pm$0.04 &\\
14944&14500$\pm$100&4.27$\pm$0.03&$-$2.29$\pm$0.09&0.51$\pm$0.04 &\\
15026&8700$\pm$100&3.08$\pm$0.05&$-$1.00$\pm$0.00&0.36$\pm$0.03 &\\
16551&14500$\pm$100&4.28$\pm$0.03&$-$2.31$\pm$0.09&0.51$\pm$0.04 &\\
15395&25700$\pm$300&5.58$\pm$0.03&$-$2.54$\pm$0.05&0.69$\pm$0.06&$\Box$\\
20919&8000$\pm$40&2.91$\pm$0.03&$-$1.00$\pm$0.00&0.37$\pm$0.03 &\\
18782&12100$\pm$100&3.78$\pm$0.03&$-$2.10$\pm$0.12&0.47$\pm$0.04 &\\
17941&24800$\pm$400&5.02$\pm$0.03&$-$1.98$\pm$0.05&0.73$\pm$0.06 &$\ast$\\
20302&19100$\pm$300&4.87$\pm$0.03&$-$1.78$\pm$0.03&0.43$\pm$0.03 &\\
26756&10430$\pm$90&3.55$\pm$0.03&$-$1.00$\pm$0.00&0.41$\pm$0.03 &\\
27181&13500$\pm$100&3.96$\pm$0.03&$-$1.98$\pm$0.09&0.38$\pm$0.03&$\triangle$\\
24849&11860$\pm$90&4.09$\pm$0.03&$-$1.65$\pm$0.09&1.02$\pm$0.03 &\\
27604&17600$\pm$200&4.60$\pm$0.02&$-$1.89$\pm$0.03&0.48$\pm$0.08 &\\
28231&26900$\pm$300&5.58$\pm$0.03&$-$1.84$\pm$0.03&0.62$\pm$0.05&$\Box$\\
26760&15600$\pm$200&4.42$\pm$0.03&$-$1.93$\pm$0.07&0.57$\pm$0.05 &\\
28554&26500$\pm$400&5.59$\pm$0.03&$-$2.33$\pm$0.05&0.70$\pm$0.06&$\Box$\\
28693&28800$\pm$400&5.56$\pm$0.03&$-$3.26$\pm$0.02&0.55$\pm$0.04 &\\
28947&22100$\pm$400&5.17$\pm$0.03&$-$1.84$\pm$0.02&0.55$\pm$0.04 &\\
4964&10740$\pm$100&3.72$\pm$0.03&$-$1.00$\pm$0.00&0.60$\pm$0.05 &\\
49317&7790$\pm$30&2.56$\pm$0.03&$-$1.00$\pm$0.00&0.17$\pm$0.01 &\\
5455&26600$\pm$300&5.63$\pm$0.03&$-$2.23$\pm$0.02&0.71$\pm$0.06&$\Box$\\
5487&20000$\pm$300&5.09$\pm$0.03&$-$1.60$\pm$0.02&0.57$\pm$0.05 &\\
5134&15200$\pm$200&4.33$\pm$0.03&$-$2.42$\pm$0.10&0.44$\pm$0.04 &\\
4672&25200$\pm$300&5.39$\pm$0.03&$-$2.04$\pm$0.03&0.53$\pm$0.04 &\\
5201&27900$\pm$300&5.53$\pm$0.03&$-$1.58$\pm$0.03&0.41$\pm$0.03 &\\
5865&27800$\pm$300&5.53$\pm$0.03&$-$3.07$\pm$0.05&0.61$\pm$0.05&$\Box$\\
7843&14100$\pm$200&4.07$\pm$0.03&$-$2.01$\pm$0.07&0.36$\pm$0.03&$\triangle$\\
6284&27200$\pm$300&5.41$\pm$0.03&$-$2.27$\pm$0.03&0.49$\pm$0.04 &\\
10257&8800$\pm$200&3.06$\pm$0.09&$-$1.00$\pm$0.00&0.31$\pm$0.03 &\\
10625&28700$\pm$300&5.67$\pm$0.03&$-$1.84$\pm$0.03&0.50$\pm$0.04 &\\
8672&30100$\pm$300&5.73$\pm$0.03&$-$2.90$\pm$0.09&0.48$\pm$0.04 &\\
10711&27700$\pm$300&5.63$\pm$0.03&$-$2.28$\pm$0.05&0.63$\pm$0.05&$\Box$\\
11609&14300$\pm$100&4.23$\pm$0.02&$-$3.04$\pm$0.16&0.51$\pm$0.04 &\\
14664&8050$\pm$40&3.02$\pm$0.03&$-$1.00$\pm$0.00&0.42$\pm$0.03 &\\
14727&10600$\pm$100&3.72$\pm$0.03&$-$1.00$\pm$0.00&0.81$\pm$0.06 &\\
35186&10800$\pm$100&3.73$\pm$0.03&$-$1.00$\pm$0.00&0.68$\pm$0.05 &\\
35662&12900$\pm$200&3.96$\pm$0.03&$-$2.15$\pm$0.16&0.43$\pm$0.03 &\\
35499&12500$\pm$100&3.96$\pm$0.03&$-$2.09$\pm$0.12&0.51$\pm$0.04 &\\
36242&12800$\pm$100&3.93$\pm$0.03&$-$2.03$\pm$0.10&0.40$\pm$0.03 &\\
36480&22400$\pm$400&5.16$\pm$0.03&$-$1.98$\pm$0.02&0.49$\pm$0.04 &\\
36502&12300$\pm$100&3.89$\pm$0.03&$-$1.92$\pm$0.10&0.46$\pm$0.04 &\\
36830&27400$\pm$300&5.63$\pm$0.03&$-$2.08$\pm$0.05&0.66$\pm$0.05&$\Box$\\
38095&14300$\pm$200&4.03$\pm$0.03&$-$1.93$\pm$0.05&0.31$\pm$0.02&$\triangle$\\
38087&27300$\pm$300&5.53$\pm$0.03&$-$2.18$\pm$0.05&0.63$\pm$0.05&$\Box$\\
32470&10620$\pm$90&3.59$\pm$0.03&$-$1.00$\pm$0.00&0.40$\pm$0.03 &\\
28695&9600$\pm$100&3.35$\pm$0.07&$-$1.00$\pm$0.00&0.40$\pm$0.04 &\\
38504&12200$\pm$100&3.90$\pm$0.03&$-$2.13$\pm$0.10&0.50$\pm$0.05 &\\
39008&30900$\pm$300&5.55$\pm$0.03&$-$2.29$\pm$0.05&0.53$\pm$0.05 &\\
38889&12700$\pm$200&3.94$\pm$0.05&$-$2.00$\pm$0.17&0.47$\pm$0.04 &\\
38963&9000$\pm$200&3.16$\pm$0.09&$-$1.00$\pm$0.00&0.32$\pm$0.04 &\\
\hline
\end{tabular*}
\end{center}
\end{table}
\clearpage

In addition, errors in the
normalization of the spectrum, imperfections of flat field/sky
background correction, etc. may produce systematic rather than
random errors, which are not represented well by the error obtained
from the fit routine.
Masses were calculated from the previously measured atmospheric parameters
in two ways. First we used the equation
\begin{equation}
\log{\frac{M}{M_{\sun}}}= const. + \log{g} + 0.4\cdot ((m - M)_V - V - V_{th})
\label{eq1}
\end{equation}
where $V_{th}$ is the brightness at the stellar surface as given by \citet{Kurucz92}.
In addition we also calculated masses through the equation
\begin{equation}
\log{\frac{M}{M_{\sun}}}= \log{\frac{g}{g_{\sun}}} - 4\cdot \log{\frac{T}{T_{\sun}}} + \log{\frac{L}{L_{\sun}}}
\label{eq2}
\end{equation}
obtained from basic relations, assuming
T$_{\sun}$=5777 K and $\log{\mathrm{g}_{\sun}}$=4.4377.  In the last
term we expressed the luminosity in terms of absolute magnitude and
bolometric correction BC through the relation
\begin{equation}
\log{\frac{L}{L_{\sun}}}= - \frac{M_\mathrm{V} + BC - 4.74}{2.5} .
\end{equation}
The bolometric correction for each target star was derived from
effective temperatures using the empirical calibration of
\citet{Flower96}. In both equations we used photometric data from
\citet{Momany02}, and we adopted an apparent distance modulus
(m-M)$_{V}$=13.17 and E$_\mathrm{B-V}$=0.04 as the mean value of the
determination of \citet{Renzini96}, \citet{Reid97,Reid98} and
\citet{Gratton97}.  Results from both equations are plotted separately
in Fig. \ref{Massesplot}. Errors on masses were derived from
propagation of errors, including the $\sqrt{3}$ factor for errors
on T$_\mathrm{eff}$ and $log{\mathrm{g}}$.
We also assumed an error of 0.1 magnitudes on the photometric quantities
used in the equations, such as distance modulus, magnitude, V$_{th}$, and
bolometric correction.

\section{Results}
\label{capresults}

\subsection{Temperatures and gravities}

\begin{figure}
\begin{center}
\resizebox{\hsize}{!}{\includegraphics{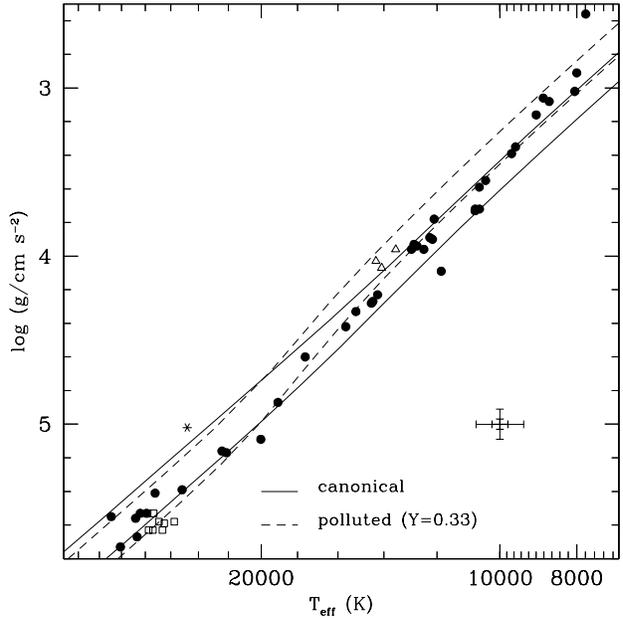}}
\caption{Temperatures and gravities of our program stars. All stars hotter than
12\,000~K or showing evidence of iron enrichment are fitted with model
atmospheres of super-solar metallicity ([M/H] = $+$0.5). Cooler stars
showing no evidence of iron enrichment are fitted with metal-poor model
atmospheres ([M/H] = $-1.5$).
The solid lines mark the zero-age (ZAHB) and terminal-age
(TAHB) loci of canonical HB tracks (Y=0.24) for [M/H] = $-$1.54 \citep[see][for
details]{Moehler03}. These loci define the region within which the HB models spend
99\% of their HB lifetime.
Dashed lines represent the ZAHB and the TAHB of helium enriched
(Y=0.33) stars. Error bars are omitted for clarity, but the typical size of
the errors as derived from the $\chi^2$ fit is indicated.
The thick error bars indicate the formal error as in Table \ref{datatable},
and the larger thin error bars indicate three times the previous value for hotter stars,
approximately 1000~K and 0.1 dex, which is probably a better estimate
of real errors (see text for details).}
\label{tgplot}
\end{center}
\end{figure}

Our results on atmospheric parameters are plotted in Fig.
\ref{tgplot}, where the position of the target stars in the
T$_\mathrm{eff}$-$log{\mathrm{g}}$ space is compared with theoretical
models. The agreement between
the measured parameters and the expectaction from the models
is good, and all stars but a few exceptions fall between
the theoretical zero-age HB (ZAHB) and the terminal-age HB (TAHB)
for normal He-content (Y=0.24) stars.

We note a group of hot stars (T$_\mathrm{eff}$$\approx$27\,000 K)
for which the measured surface gravities seem to be systematically
higher than expected for normal He content stars.
Another group of three cooler stars (T$_\mathrm{eff}$$\approx$14\,000 K)
shows gravities that are too low and lies above the TAHB.
Though these differences are marginal in the
T$_\mathrm{eff}$-$log{\mathrm{g}}$ plot, they appear more significant
when we consider the masses (see Fig.~\ref{Massesplot}).   
It is worth noticing in Fig \ref{tgplot} that they are all reproduced
better by models with enhanced helium abundance (Y=0.33).

There is an additional anomalous star at
T$_\mathrm{eff}$$\approx$25\,000 K, which has a significantly smaller
surface gravity than expected. This star
(marked as asterisk in all the figures in this paper)
also has an anomalous position in the color-magnitude diagram
and will be discussed later.
All the anomalous stars have radial velocities compatible with the
radial velocity of \object{NGC6752}, as can be seen in Table 1
of Paper I. Therefore they can be considered probable cluster
members, but it must be noted that radial velocity alone is not
an unambiguous proof of membership in this cluster. In fact, at the
galactic coordinates of \object{NGC6752} ({\it l}=336\degr ,
{\it b}= $-$26\degr), a certain amount of field contamination is expected,
and its radial velocity is unremarkable compared to field disk stars
(-27.9 km s$^{-1}$, \citealt{Harris96}).

\begin{figure}
\begin{center}
\resizebox{\hsize}{!}{\includegraphics{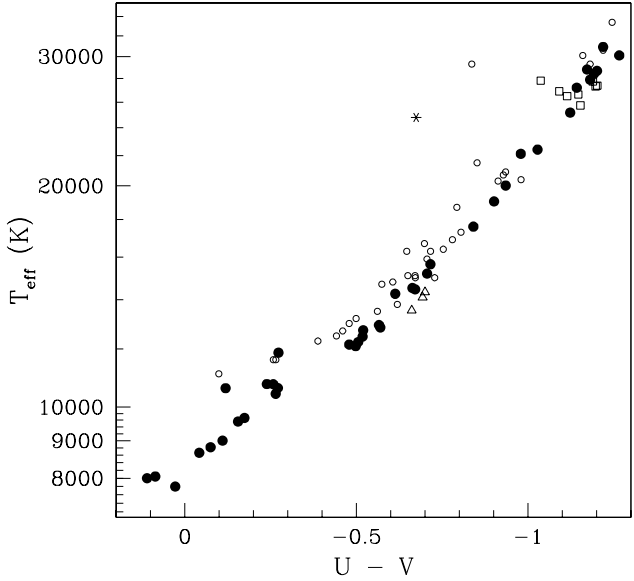}}
\caption{Measured effective temperatures as a function of colors derived
in this work (full circles, and triangles and squares follow the notation in previous plots)
and in \citet{Moehler00} (small empty circles).
The adopted data from \citet{Moehler00} are the ones obtained with
metal-poor model atmospheres (their Table 2) for stars with T$_\mathrm{eff}\leq
11\,000$~K and with metal-rich model atmospheres (their Table 5) for hotter stars.}
\label{TeffM00}
\end{center}
\end{figure}

We compared the derived effective temperatures with those measured in
\citet{Moehler00} for HB stars in the same cluster.  The temperatures
of both data sets are plotted as a function of color in Fig.
\ref{TeffM00}. The results agree very well.
There is a small offset of our data toward lower temperatures, but it cannot
be considered significant because it is much smaller ($\leq 500$~K)
than formal errors.  We also have nine stars in common
with \citet{Moehler00}. In Fig. \ref{TeffM00} we plot the
differences between the two works in derived $\log{\mathrm{g}}$ and
masses. Within the errors, the results of the two investigations are consistent,
as was expected due to the similarities between the two works:
data reductions are almost identical, and measurement procedures and
model atmospheres are the same.

\subsection{Masses}

In Fig. \ref{Massesplot} we plot the masses
we derived for our target stars as a function of effective
temperatures, with both Eqs. \ref{eq1} and \ref{eq2}.
Here the problems and the differences between observed
and expected values become evident.
We note that both equations give very similar results, and the
differences are negligible in all but the temperature range
12\,000-15\,000 K. In this interval the masses calculated with Eq.
\ref{eq2} show better agreement with theoretical expectations because they
are on average 0.05 M$_{\sun}$ higher, but it is not enough to correct the
general tendency of an underestimation of the derived masses for
T$_\mathrm{eff}\leq$15\,000 K.
We discuss this problem in more details in subsequent sections.

\begin{figure}
\begin{center}
\resizebox{\hsize}{!}{\includegraphics{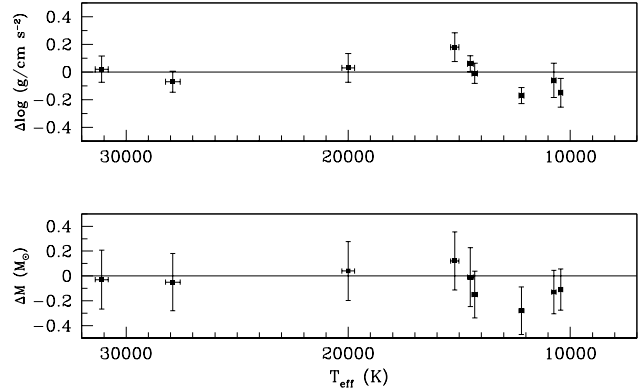}}
\caption{Differences in surface gravities and derived masses between this work and
\citet{Moehler00} for the nine stars in common. The errors are the quadratic sum of
the ones for each set of data. The difference is in the sense (ours) - (M00).}
\label{bMM00}
\end{center}
\end{figure}

Here we note that, although the plotted points in this temperature range
are closer to the
theoretical line in the lower panel of Fig. \ref{Massesplot}, they
are not scattered around it, but systematically below it, as in
the upper panel. This indicates that the problems concerning derived
masses cannot be ascribed only to either the theoretical V$_{th}$ values
used in Eq. \ref{eq1} or to empirical BCs used in Eq. \ref{eq2}, because both
equations lead to the same underestimation independently.

The problem cannot  be attributed to the $V$ magnitudes from \citet{Momany02}
that  we use to derive the masses.
When calibrating photometric data, we all refer to the  standard $UBV$ Johnson
filters and calibrate with respect to them. However, the study by
\citet{Momany03} basically shows that no 2
filter-systems are the same (see their Fig.2) and that small
differences in the bandpass of the employed filters (at any telescope)
may persist. Momany et al. also show that such ``small''
differences can eventually give rise to un-physical features.
and we are aware  that the WFI@2.2m $V$  filter has a red
cutoff at $\lambda\sim5900\AA$ with respect to the standard $V$
filter. Moreover, the WFI@2.2m instrumental $v$ magnitudes were
calibrated with their respective ($b-v$) colors, thus employing  the
WFI@2.2m $B$ filter with a redder effective central wavelength.
Nevertheless, it remains quite hard  to explain the derived low masses
in  terms of photometric  calibration uncertainties. Indeed, to derive
masses that match theoretical expectations, we estimate that
the M$_V$ magnitudes need to be increased by  $\sim0.4$  in  the
$12\,000-15\,000$ K range, and decreased by a similar amount for the hot
outliers. Clearly, this discrepancy is too large to be attributed to
photometric calibration uncertainties.

Because of the lack of noticeable differences between the panels in Fig.
\ref{Massesplot}, we discuss the results without distinguishing
between the two methods to derive masses. Instead, we find it
appropriate to divide the sample to three different
ranges of temperatures.

\subsubsection{Masses: cool stars (T$_\mathrm{eff}\leq$12\,000 K)}
\label{massescool}

The derived masses of stars cooler than 10\,000 K are systematically too low.
All of these stars are offset from the HB towards lower gravities in Fig.~\ref{tgplot}.
\citet{Moehler03} observed the same effect for HB stars in M13 and
excluded these stars from their analysis.
It is worth noticing that \citet{Moehler00} find masses in this range
of temperatures in good agreement with the expected value for stars in the
same cluster, but \citet{Moehler06} in \object{NGC6388} find masses that
are too low, as we do with the same instruments and set-up as in the present work (UT4+FORS2 and grism B600).
They consider that atmospheric parameters for these stars
are probably not trustworthy, because of the presence of metallic lines not
considered in the fitting procedure and not clearly observable due to
low resolution.  In the case of \object{NGC6752}, metal lines offer
less explanation because this cluster is more than a factor of 10 more
metal-poor than \object{NGC6388}. However because of the coincidence observed,
it can be hypothesized that, whatever the cause,
it could somehow be enhanced by the instrumentation and set-up used.
\citet{Moehler03} find too low mass for the star they observed
in this temperature range, in a cluster of similar metallicity (\object{M3})
but with different instrumentation.
This excludes the possibility that the underestimation was only due to
instrumental effect with this set-up,
and the problem of the masses too low for T$_\mathrm{eff}\leq$10\,000 K
remains completely open.
The calculated masses for stars between 10\,000 and 12\,000 K
better agree with theoretical values although the scatter is quite high.

\subsubsection{Masses: intermediate temperature stars
(12\,000$\leq$T$_\mathrm{eff}\leq$15\,000 K)}
\label{massesintermediate}

\begin{figure}
\begin{center}
\resizebox{\hsize}{!}{\includegraphics{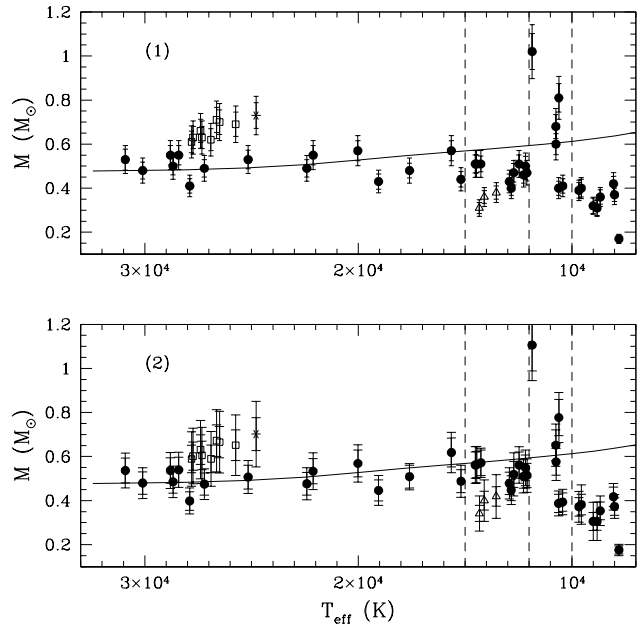}}
\caption{Estimated masses of target stars as a function of effective temperature.
Different symbols are used to distinguish between groups of stars that behave
differently, as indicated in the text. Also the theoretical
HB \citep{Moehler03} is indicated. Error bars on temperatures are omitted for clarity,
but are within an order of magnitude of full points. Errors on masses are obtained
from the propagation of errors on temperature and gravities, this error bars referring
to the formal errors given in Table \ref{datatable}, and larger ones assuming errors
of 1000~K and 0.1 dex respectively, which are probably more realistic estimates
of real errors. Vertical dashed lines divide the temperature intervals analyzed
in the text. {\it Upper panel}: masses derived from Eq. \ref{eq1}.
{\it Lower panel}: masses derived from Eq. \ref{eq2}.}
\label{Massesplot}
\end{center}
\end{figure}

For higher temperatures, as the radiative levitation sets in, masses
are constantly underestimated. 
This is a well-known result that is clearly present in our data.
In fact, even if the errors on the masses of the single
stars are high,
Fig. \ref{Massesplot} clearly shows that our derived  masses in the
12\,000$\leq$T$_\mathrm{eff}\leq$15\,000 K interval
are systematically lower than predicted.
Our results are also in agreement with the ones of
\citet{Moehler00}, and the difference that we find between observed
and expected masses is of the same order of magnitude as in their
work. We then confirm their conclusions that the use of models with
higher metallicity in the fitting procedure, to account for radiative
levitation, can only partially reduce the discrepancy.

There is a group of three stars located at
$13\,000<$T$_\mathrm{eff}<14\,000$ K with particularly low masses
and low gravities. These stars are indicated in all of the figures as empty
triangles. These stars are also separated well in the color-magnitude
diagram (Fig. \ref{cmd}), as they are
systematically brighter than the others. It must be emphasized that,
although we relied on photometric data to derive masses, the
difference in masses cannot be ascribed to the difference in
magnitude, since they go in opposite directions: all other parameters
being the same, brighter stars should show higher masses, and not
lower, as can be deduced by Eq. \ref{eq1}.
These anomalous masses are probably due to the very low surface
gravities, low enough to give underestimated masses, even if  the stars
are brighter than their neighbors in the color-magnitude diagram.
Therefore, for some reason, these stars have different luminosities {\it and}
a different spectrum than other HB stars of similar temperatures.
In this temperature range, there is another star (\#18782) brighter
than the average HB in Fig. \ref{cmd} and outside (above) the
theoretical tracks in Fig. \ref{tgplot}.  Nevertheless its
calculated mass is not noticeably low with respect to other stars with
similar temperatures, so we did not point it out with a different
symbol. We do take it into account in the later discussion.

\begin{figure}
\begin{center}
\resizebox{\hsize}{!}{\includegraphics{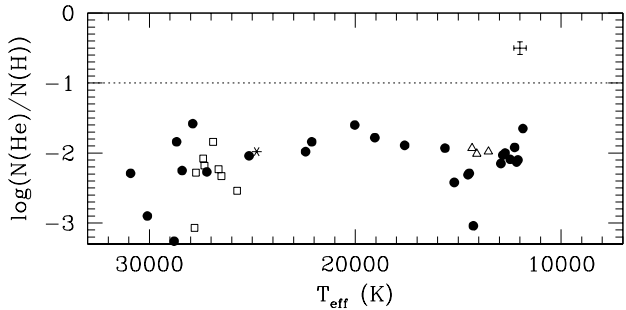}}
\caption{Derived Helium abundances as a function of effective temperatures for
target stars. Different
symbols are used to distinguish between groups of stars that behave
differently, as indicated in the text. The dotted line indicates
solar value, and the typical errorbars (0.09 dex in log(N((He)/N(H)) and 300 K in temperature)
are indicated in upper right. For stars cooler than the Grundhal jump
the abundance was kept fixed to solar value in the fitting procedure,
and they are omitted in the plot.}
\label{Heplot}
\end{center}
\end{figure}

\subsubsection{Masses: hot stars (T$_\mathrm{eff}\geq$15\,000 K)}
\label{masseshot}

For increasing temperatures, the measured masses again generally
agree with theoretical values, as also found by \citet{Moehler00} in the same
cluster. Nevertheless, also among hotter stars there is a
group of stars that behave differently from the general trend. These
objects are indicated as empty squares in all the plots, and for them
the derived masses are systematically higher than expected. As shown in 
Fig. \ref{cmd}, they
are fainter than the bulk of HB stars at the same temperature. Again,
Eq. \ref{eq1} would imply lower masses for fainter stars,
so this is not caused by how we derived our masses:
either these objects are intrinsically more massive or they are
both photometrically and spectroscopically different from other stars at the same temperature.

\subsection{Helium abundances}
\label{Helium}

The derived He abundances are plotted in Fig. \ref{Heplot} as a
function of effective temperatures, where only stars hotter than
Grundhal jump are considered because, for cooler ones, He abundance was
not measurable, hence kept fixed to solar value.  All stars show helium
depletion to a different extent, but no trend with temperature is visible.

The three stars indicated as triangles, which show particularly low
gravities and/or masses, on average also show a higher helium
abundance than do other stars at the same temperature,
though they agree with values for stars at different
temperatures - from cooler ones up to 20\,000~K.
On the other hand, no clear trend is visible for hot stars with anomalous
measured masses, which show helium abundances in agreement
with other hot stars.

\subsection{Star \#17941}
\label{stars9a}

Star \#17941 is very peculiar and its measured atmospheric parameters
are highly uncertain. This object is indicated with an asterisk in all
the figures.

Its radial velocity, as measured in Paper I, indicates that it is a probable
cluster member. In the color-magnitude diagram it is located below the HB at
fainter magnitudes and/or redder color.
Its B-V color indicates an effective temperature around 14\,000 K, which is also
confirmed by its intermediate-resolution spectrum (Paper I), which shows the typical
features of an HB star with this temperature, in particular, a high quantity of metallic
lines due to radiative levitation. Actually, its spectrum is very similar to the
ones of other stars in this temperature range. On the contrary, from the measurements on
low-resolution spectra, it comes out to be a much hotter HB star (24800 K). Also
the surface gravity is peculiar, being particularly low. Even more puzzling,
and at variance with the behavior of all other peculiar stars already analyzed, we
find a {\it higher} mass than expected even if its surface gravity is {\it lower}
than canonical models.

We analyzed the data reduction of these spectra in details,
trying different solutions for spectra reduction and extraction, with
no significant changes in the derived parameters. We found no evidence
of light contamination by a nearby star, both when analyzing the slit images and
in the field images acquired before exposures. We did not find any
indication of any companion, not even among the spectral features
observed at intermediate resolution, although this could not be
indicative because the resolution itself is not very high (1.2 \AA),
and the great quantity of metallic lines could easily hide fainter
features from a companion.

\subsection{Reddening}
\label{ebv}

\begin{figure}
\begin{center}
\resizebox{\hsize}{!}{\includegraphics{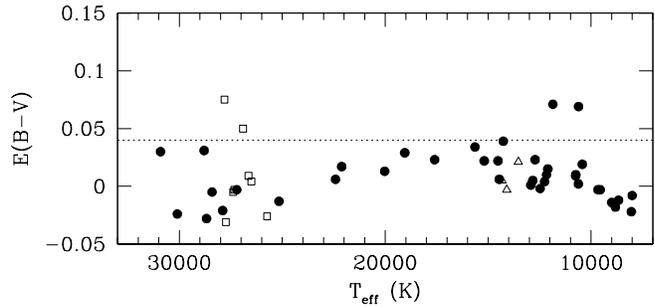}}
\caption{Estimated E$_\mathrm{(B-V)}$ reddening values for target stars. The dotted line
indicates the value E$_\mathrm{(B-V)}$=0.04, which is the mean value of the determination
of \citet{Renzini96}, \citet{Reid97,Reid98}, and \citet{Gratton97}.}
\label{figebv}
\end{center}
\end{figure}

With the derived effective temperatures and gravities, we estimated the
color excess E(B-V) that could be deduced from these atmospheric
parameters. We used the Kurucz ATLAS9 \citep{Kurucz93} photometry
table for the same metallicity as was used in fitting procedure, and for each
star we interpolated the B-V values in the grid to get the expected
values. Then we calculated the color eccess E(B-V) by comparing this
theoretical color with the observed one from \citet{Momany02}
photometry. The results are plotted in Fig. \ref{figebv}.

The computed reddening is constantly too low compared to
literature values. In fact the mean value of \citet{Renzini96},
\citet{Reid97,Reid98}, and \citet{Gratton97} that we also used in
Eq. \ref{eq1} is E(B-V)=0.04, and it is in fair agreement with
the more recent measurement by \citet{Gratton05}. We obtain E(B-V)=0.016
as the mean value for our stars, with no particular trend with temperature.

It should be noted that the WFI@2.2m bandpasses differ significantly 
from the standard Johnson-Cousins ones \citep{Momany02},
and an error of 0.02 magnitudes in color cannot be excluded.
However, \citet{Gratton05} use the same photometric data and obtain
reddening estimates in perfect agreement with previous works among
turn-off stars in the same cluster.
The main difference is that they estimated temperatures fitting
H$_{\alpha}$ profiles in high-resolution spectra, a totally different
procedure from the one adopted here.  Therefore, we are
forced to conclude that the offset observed here could probably point
to a systematic underestimate of the temperature along the HB.  Of
course, the problem could reside in the theoretical colors
themselves. In fact, empirical calibrations of T$_\mathrm{eff}$-color
relations show that synthetic colors from model atmospheres
calculated with ATLAS9 have difficulty properly reproducing
observations \citep{Ramirez05,Sekiguchi00}, a problem also pointed out
in the theoretical investigation by \citet{Castelli97}.
Unfortunately, all the cited empirical calibrations are limited in the
cool temperature range (spectral type G-K) of main sequence or giant stars,
and there is a general lack of HB color-temperature, model-independent
calibrations in the literature. In conclusion, it is not possible to guess to what extent
ATLAS9 models could lead to a systematic error in temperature.

The stars that show peculiar masses show
no particular behavior in Fig. \ref{figebv}. Since reddening
estimates are very sensitive to temperature but should hardly change
for small variations of surface gravity, this can indicate that their
derived temperature are not peculiar, and their strange behavior is only
due to an unusual surface gravity (either real or as a consequence of
other effects unaccounted for).


\section{Discussion}
\label{discussion}

We assumed LTE in our model spectra, as usual in this kind of
analysis, but it is becoming evident that non-LTE effects could affect
the results for hotter stars and lead to erroneous atmospheric
parameters, as pointed out by \citet{Przybilla06}. In their
preliminary results they show that a hybrid NLTE analysis can provide
a much better fit of H and He lines, but the differences in
atmospheric parameters relative to results assuming LTE are
only slightly larger than our real uncertainties (on the order of 1000 K
in T$_{\mathrm{eff}}$ and 0.1 dex in $log{\mathrm{g}}$).
Considering that errors in Table \ref{datatable} are underestimated by
a factor of at least 2--4, as already discussed, the influence of neglecting
NLTE effects is relatively small. Even more important, it should
affect all stars in the same way at a given temperature
(NLTE analysis provides higher temperatures and gravities than
the LTE one), so it should not cause any systematic difference
for stars at approximately the same temperature.
However, adding 1000 K in T$_{\mathrm{eff}}$ and 0.1 dex in $log{\mathrm{g}}$
strongly increases the derived masses as calculated from Eq.
\ref{eq1}. The change in temperature is reflected in a decrease of
$\approx$0.1 mag for V$_{th}$, which is quite insensitive to gravity.
Masses are then higher by about 0.2~M$_{\sun}$ in the high temperature
range.  The effect on masses derived from Eq. \ref{eq2} is much
smaller, approximately half the previous one, but again in the sense
of increasing masses. While this would alleviate our problems
between 12\,000~K and 15\,000~K, there is no reason to assume that NLTE
effects would be limited to this temperature range.

\begin{figure}
\begin{center}
\resizebox{\hsize}{!}{\includegraphics{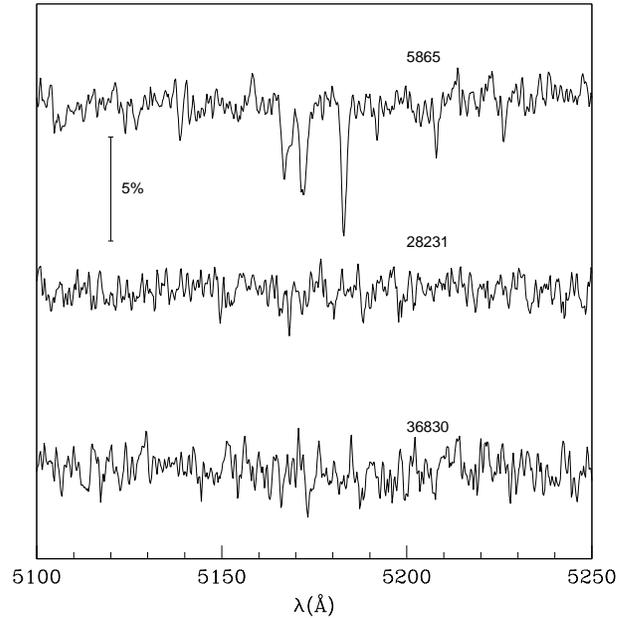}}
\caption{A section of the spectrum of star \# 5865 where the MgI triplet is
clearly visible. The spectra of two other peculiar hot stars are plotted for comparison.
The spectra have been obtained for each star by summing the high-resolution ones used in Paper I,
after correcting for RV shifts.}
\label{Figcompanion}
\end{center}
\end{figure}

We also assumed solar abundance ratios in the models. This is
necessarily a crude approximation in order to take into account the
enhanced abundancies of heavy elements due to radiative
levitation. Detailed studies both in the proximity of the Grundhal
Jump \citep{Behr99,Behr03} and among hotter HB stars
\citep{Edelmann01} show that the atmospheric metal abundances in the
presence of radiative levitation do not usually follow solar ratios:
the elements are affected to different extents and some are even not
enhanced at all.
The pattern of abundance ratios and its effect on the atmospheric
structure are still poorly understood.
The approximation of solar abundance ratio could introduce a
systematic effect in parameter determination to an unknown extent, and
in principle stars with different abundance patterns (due to different
formation history) could behave differently in our analysis. Eitherway,
there are indications \citep{Edelmann06} that, at least for EHB stars,
the surface abundance pattern could almost be independent of stellar
parameters, so any systematic effects, if relevant, should be the same in a
wide range of temperatures and gravities.
This might not be the case for cooler stars. In fact, \citet{Pace06} find a clear
trend toward increasing heavy elements abundance with effective temperatures across the
Grundhal jump ($10\,000 \leq \mathrm{T_{eff}} \leq 14\,000~K$). Therefore, at least
in this temperature range, the deviations from solar abundance ratio (and then
the differences with respect to the models) should be increasing with effective temperature.
This could help for explaining our results, but the extent of the consequences on
derived parameters is unknown, and in Fig. \ref{Massesplot} a clear trend with
temperature in this range is not visible.

Our results for the three peculiar stars at 14\,000 K appear to
be compatible with the hypothesis of
helium-enrichment (as shown in Fig. \ref{tgplot}).  However, in this case the low
surface gravity is directly linked to the higher luminosity as shown
by \citet{Catelan05} and \citet{Sweigart98}. The masses derived from
effective temperatures, surface gravities, and luminosities should
therefore be well above 0.5~M$_{\sun}$.  The low masses derived here
suggest that the low gravities are at least partly spurious.
The same reasoning seems to exclude the hypothesis that these stars are an
evolving or evolved object, as could be guessed looking at their position
in the color-magnitude diagram. In fact, in this case again, the higher
luminosities would compensate
for the lower gravities, and the masses should agree with expectations.
This is what we observe for star \#18782 at U-V$\approx$-0.5, which
shows both brighter luminosity and lower gravity than other
stars, but this is not peculiar for its calculated mass. Therefore we
suppose that helium-enrichment or evolutionary effects can be
considered good hypothesies that explain our observations for this star.

It must be noted
that the measured helium abundance cannot be properly disentangled
from measured gravities because both have been measured with the same
fitting procedure. Helium abundances are mainly measured from He lines, but
as investigated by \citet{Moehler06},
varying the helium abundance directly changes the width of Balmer
lines, in the sense that higher helium abundances should
produce broader lines. This rules out the possibility that low
gravities are produced by an unnoticed helium enrichment, but
indicates that underestimating gravity and overestimating helium abundance
are two correlated effects. That these stars are
systematically brighter prevents concluding that the anomalous
parameter comes only from this weakness in the fitting procedure.
These stars must be intrinsically different from fainter ones.

Probably the best explanation is that our results
indicate enhanced stellar winds. In fact \citet{Vink02} indicate
that a non negligible wind should be expected for HB stars. They also
show how it can change Balmer line profiles in such a way as to mimic a
lower surface gravity. There is still a lack of detailed models that
take the presence of wind into account, making it necessarily
unaccounted for in our fitting procedures. Then, the derived low
surface gravities could result from this fault.
The wind scenario is corroborated by the anomalous stars being brighter,
so an enhanced stellar wind could be expected.
Moreover, a wind could possibly counteract the diffusion, which reduces the
atmospheric helium abundance observed in subdwarf B stars to sub-solar
values and thus justifies the higher helium abundances we measure.

It is harder to find a good explanation for our results on the eight hot
peculiar stars indicated as open squares in all plots. They were first spotted
in Fig. \ref{Massesplot} for their abnormally high masses, and then
individuated in the color magnitude diagram as being fainter and/or redder
than other stars observed in the same color-magnitude range.

In this case we can assume that spectroscopically derived masses
and photometric properties are independent. It is not generally true because
magnitudes were used in the mass calculation, but as we saw, high masses cannot be
ascribed to fainter magnitudes, and then the two effect cannot be related. The
cause of peculiar masses must be other than photometric data, and
we would expect these stars to be randomly distributed among the EHB population.
Under this assumption, we performed a two-dimensional Kolmogorov-Smirnov (KS) test
to explore the possibility that these stars are also photometrically
peculiar.
The two-dimensional version of the KS-test is not mathematically well-defined,
but we adopted the algorithm of \citet{Peacock83}, which gives reliable results
for all the cases of practical interest.
We compared the distribution of the eight stars in the color-magnitude
diagram with the whole population of stars in the range -1.25$\leq$(U-V)$\leq$-1 and
17$\leq$V$\leq$17.5. The test reveals that there is a 8.4\% probability that the
analyzed sample is randomly drawn
from the test distribution. This result is clearly not conclusive, but indicates
that with high probability (91.6\%) the eight hot peculiar stars indeed show
photometric peculiarities and could belong to a sub-population of stars
different from the brighter and bluer bulk of EHB star population.
It can be excluded that this result stems from a selection effect
in the target selection, because the same test, when applied to all 13
spectroscopically observed stars in this temperature range, gives a 25\%
probability of this sample being drawn from the test distribution.

For the eight hot peculiar stars, the helium
enrichment hypothesis is a possible explanation.
In fact, in the T$_\mathrm{eff}$-$\log{\mathrm{g}}$ plot these stars
agree better with polluted models and,
according to calculations by \citet{Sweigart98}, an increased helium abundance
in the high temperature regime implies higher gravity.
However, in this case their overly high masses again put this explanation
in question, because higher gravities and lower magnitudes
would balance out and lead to normal derived masses.

As these stars are redder than their neighbors in the
color-magnitude diagram, they might be binaries. In this case,
contamination by the light of a cool companion could affect the
parameter determination, and additional light would increase any
derived mass.
We therefore analyzed the spectra looking for signatures of the
presence of cool companions, both in the low-resolution spectra used
in this work and the higher resolution ones from Paper I, focusing on
features typical of G-K type stars, such as the G-band and the MgIb
triplet. We found no evidence of companions, except
for star \#5865, which shows a strong MgI triplet, unusual for such a
hot star (see Fig. \ref{Figcompanion}).
Although this could indicate
a binary system, it is just an isolated case, and the failure to
individuate these features in all the other seven stars puts the
hypothesis of binarity for these peculiar hot stars in doubt. In addition
we did not find any indication in Paper~I of radial velocity variations for
any of these stars.

In principle, we cannot exclude that their masses are actually
higher, but it appears to be quite improbable and hard to explain.
It is simpler to think that the gravities are overestimated by fitting
procedures, but even this conclusion has important
implications. As a consequence, we must deduce that models
are unable to describe these objects and their spectra properly.
Some difference between the model and the real star should be responsible
for unreliably measured masses. But
the same models give results in perfect agreement with theoretical
expectations for ``normal'' stars, therefore the ``peculiar'' ones must be
intrinsically different from the main population. This conclusion is
strongly corroborated by the color-magnitude diagram, where peculiar stars seem to be
fainter and/or redder than the others.

We point out that the hot peculiar stars lie in concomitance of the Momany
Jump \citep{Momany02}, a feature discovered first in
\object{NGC6752} and then also individuated in other clusters \citep{Momany04}. The
Jump is located around T$_\mathrm{eff}$=23\,000 K and is characterized
by a sudden increase in luminosity of the HB stars.  As suggested by the
author of the discovery paper, the presence of the jump could be due
to deep changes in atmospheric structure at this critical temperature,
with the onset of radiative levitation and changes in chemical
abundances.
We find no good explanation for our results in this direction, since
we would expect brighter stars (and not fainter) to be deviating with
respect to the models, but it is a coincidence that probably should be
investigated further.


\section{Conclusions}
\label{conclusions}

We measured atmospheric parameters (T$_\mathrm{eff}$,
$\log{\mathrm{g}}$, $\log{\frac{N(He)}{N(H)}}$) and derived masses for
51 HB stars in \object{NGC6752}.  Our results
agree in general with previous studies, but not always with theoretical
expectations. Like previous works in literature, we find masses that are
too low for cool (T$_\mathrm{eff}\leq 10\,000$~K) and
intermediate-temperature (12\,000$\leq$T$_\mathrm{eff}\leq 15\,000$~K)
stars.  We find two groups of stars deviating with respect to
the general behavior of the sample. The first group at 14,000\,K
shows masses that are too low, for which we consider a weak stellar wind
(unaccounted for in the model spectra) as the
most probable explanation. The second group at about 27\,000~K shows
too high masses, for which we found no good explanation. These stars
most probably (91.6\%) do not belong to the general distribution of
EHB stars in the color-magnitude diagram. We conclude that
atmospheric models, successful in reproducing the other EHB stars (for
which we obtain masses in agreement with expectations), are inadequate
for these peculiar ones.


\begin{acknowledgements}

CMB acknowledges Universidad de Chile graduate fellowship support from
programs MECE Educaci\'on Superior UCH0118 and Fundaci\'on Andes C-13798.
We want to thank the staff at the La Silla Paranal Observatory for their
support during our observations.
\end{acknowledgements}


\bibliographystyle{aa}
\bibliography{biblio}

\end{document}